\documentclass[12pt]{article}
\usepackage{amsmath,amsfonts,amssymb,graphicx,psfrag,rotating}
\usepackage[colorlinks=true,urlcolor=blue,citecolor=black,linkcolor=blue]{hyperref}
\usepackage{setspace}
\usepackage{dcolumn}
\usepackage[superscript,biblabel]{cite}

\usepackage{comment}
\usepackage{multirow}
\usepackage{lscape}
\usepackage{bbm}
\usepackage{scalefnt}
\usepackage[autolanguage]{numprint}
\usepackage{booktabs}
\usepackage{bm}
\usepackage{threeparttable}
\usepackage{rotating}
\usepackage{color}
\usepackage{setspace}
\usepackage{sectsty}

\usepackage{caption}
\usepackage{subcaption}

\usepackage{color}
\definecolor{qz}{RGB}{255,0,0}

\usepackage{lmodern}
\fontfamily{lmtt}\selectfont
\usepackage[T1]{fontenc}

\usepackage{float}

\usepackage{authblk}

\usepackage[top=1in, left=1in, right=1in,
bottom=1in]{geometry}

\makeatletter
\renewcommand{\section}{\@startsection{section}{1}{0em}{\baselineskip}{0.5\baselineskip}{\large\bfseries\large}}
\renewcommand{\subsection}{\@startsection{subsection}{0}{0em}{\baselineskip}{0.5\baselineskip}{\normalfont\bfseries\normalsize}}
\makeatletter

\newcolumntype{.}{D{.}{.}{-1}}
\newcolumntype{d}[1]{D{.}{.}{#1}}

\begin{document}

\title{Evaluating A Key Instrumental Variable Assumption Using Randomization Tests}

\author[1]{Zach Branson}
\author[2]{Luke Keele}
\affil[1]{Carnegie Mellon University}
\affil[2]{University of Pennsylvania}
\date{\today}

\maketitle

\pagestyle{plain}

\doublespacing

\begin{abstract}
\noindent
Instrumental variable (IV) analyses are becoming common in health services research and epidemiology. Most IV analyses use naturally occurring instruments, such as distance to a hospital. In these analyses, investigators must assume the instrument is as-if randomly assigned. This assumption cannot be tested directly, but it can be falsified. Most falsification tests in the literature compare relative prevalence or bias in observed covariates between the instrument and the exposure. These tests require investigators to make a covariate-by-covariate judgment about the validity of the IV design. Often, only some of the covariates are well-balanced, making it unclear if as-if randomization can be assumed for the instrument across all covariates. We propose an alternative falsification test that compares IV balance or bias to the balance or bias that would have been produced under randomization. A key advantage of our test is that it allows for global balance measures as well as easily interpretable graphical comparisons. Furthermore, our test does not rely on any parametric assumptions and can be used to validly assess if the instrument is significantly closer to being as-if randomized than the exposure. We demonstrate our approach on a recent IV application that uses bed availability in the intensive care unit (ICU) as an instrument for admission to the ICU.
 \end{abstract}

\clearpage

\section{Introduction} \label{sec:intro}

While instrumental variable (IV) methods have a long history in economics, they have become increasingly popular in epidemiology and health services research. While IV methods can be used to analyze randomized trials with noncompliance \cite{imbens1997bayesian,dunn2005estimating}, investigators commonly seek to exploit natural experiments by identifying instruments that act as haphazard nudges that induce exposure to a treatment. Instruments are valuable because, subject to a set of assumptions, they provide valid causal inferences in the presence of unobserved confounders \cite{angrist1996identification,baiocchi2014instrumental}. However, some of these assumptions cannot be tested with observed data, and recent research has focused on developing guidelines, diagnostics, and falsification tests for IV analyses.\cite{baiocchi2012near,swanson2013commentary,Yang:2013,glymour2012credible,kang2013causal,pizer2016falsification,keeleexrest2018}

One strand of this literature has focused on developing methods for assessing whether the assumption of no unmeasured confounding between the IV and the outcome is plausible. This assumption requires careful assessment, because IVs based on natural experiments are not assigned using actual random assignment. Instead, investigators seek to find natural circumstances that might produce as-if random assignment of the IV. If the IV behaves in an as-if random fashion, then baseline covariates should not differ significantly across levels of the IV. Consistent with this logic, analysts often compute the observed covariate balance across levels of the IV and compare it with balance across levels of the treatment \cite{brookhart2007preference,baiocchi2014instrumental}. If covariate balance across levels of the IV is not at least better than covariate balance across the treatment, this is one indication that the no unmeasured confounding assumption is untenable. Alternatively, bias in covariates from an IV can be compared to bias from an intention-to-treat analysis \cite{baiocchi2014instrumental}. Jackson and Swanson\cite{jackson2015toward} outlined a method of graphical display for easy bias comparisons, and Davies et al.\cite{davies2015commentary,davies2017compare} suggest using formal statistical tests to compare the bias associated with the IV to the bias associated with the treatment. More recently, Zhao and Small \cite{zhao2018} developed plots for bias that account for whether observed confounders are related to the outcome.
   
There are, however, key limitations to extant methods. First, current tests only provide covariate-by-covariate results, often making it hard to discern an overall pattern in whether the data favor the IV relative to the treatment. Second, all of the above methods rely on parametric assumptions. Third, while these tests compare the IV to the treatment, they do not compare balance on the IV and the treatment to an overall standard. As an alternative, we develop a new diagnostic test for assessing IV--outcome confounding. Our test compares IV and treatment balance or bias to a common benchmark: the balance or bias we would expect if either the IV or treatment had been randomized. The results from our test can be plotted for each covariate, but our method also allows for a global test. Moreover, our test is nonparametric, and thus relies on minimal assumptions.

In Section \ref{sec:notation}, we outline notation and the necessary assumptions for an IV analysis. In Section \ref{s:randomizationTest}, we propose our falsification test for IV--outcome confounding, and in Section \ref{s:application} we apply our diagnostics to a recent IV application that uses bed availability in the intensive care unit (ICU) as an instrument for admission to the ICU. In Section \ref{s:conclusion}, we conclude.

\section{Notation and IV assumptions} \label{sec:notation}

Let $D$ be the binary treatment exposure, $Y$ be the outcome, $Z$ be a binary instrument, and $\bm{X} = (X_1,\ldots,X_K)$ be a set of covariates measured before $Z$ and $D$ are assigned. For $z = 0,1$ and $d = 0,1$, let $Y{(z,d)}$ be the potential outcome of $Y$ when an intervention sets $Z=z$ and $D=d$, and let $D{(z)}$ be the potential outcome of $D$ when $Z = z$. We also assume there are $N$ units, indexed by $i=1,\dots,N$, though we do not subscript by $i$ in this section. The observed values $(Y, D, Z)$ are related to the potential outcomes through the following equations:
\[
D = D{(Z)} = Z D{(1)}  + (1 - Z) D{(0)},
\]
\[
Y  = Y{(Z,D)} =  Y{\left(Z, D{(Z)}\right)} = Z Y{\left(1,D{(1)} \right)} + (1 - Z) Y{\left(0,D{(0)} \right)}.
\]
This notation implicitly assumes the stable unit treatment value assumption (SUTVA) \cite{rubin1980randomization}. SUTVA implies that 1) the levels of $D$ (1 and 0) adequately represent all versions of the treatment, and 2) each subject's outcomes are not affected by other subjects' exposures.

In an IV analysis, investigators assume three ``core assumptions'' hold\cite{baiocchi2014instrumental}:
\begin{description}
\item[(1) Relevance:] $\mathbb{E}[D|\bm X,Z=1] \ne \mathbb{E}[D|\bm X,Z=0]$.
\item[(2) Effective random assignment:] $Z$ is independent of $D(z)$
  and $Y(z,d)$ given $\bm X$ for all $z = 0,1$ and $d = 0,1$. This
  implies that there are no unmeasured confounders for $Y$ and $Z$ conditional
  on $\bm X$.
\item[(3) Exclusion restriction:] $Y(z,d) = Y(z',d)$ for any $z$,
  $z'$, and $d$. This means that the potential outcomes only depend on the instrument $Z$ through its effect on the exposure $D$. This assumption allows us to rewrite the potential outcomes as $Y(0,d) = Y(1,d) = Y(d)$.
\end{description}

The methods we develop in Section \ref{s:randomizationTest} are designed to shed light on the plausibility of Assumption (2). In randomized experiments with noncompliance, Assumption (2) holds by design. However, for IVs based on natural experiments, investigators must assume that the IV is as-if randomly assigned to the units under study. Assumption (2) cannot be directly tested, but observed data can be used to falsify it. That is, an IV that is as-if randomly assigned should be independent of observed covariates. If we find that an IV is independent of observables, that tells us little about whether an IV is independent of unobservables; however, if an IV is strongly associated with observables, that suggests that Assumption (2) may not hold. Next, we outline a test to probe Assumption (2).

%However, the methods in Section \ref{s:randomizationTest} can provide a more nuanced assessment of Assumption (2) beyond a binary plausible-or-not-plausible test for this assumption. In particular, even if the instrument is found to not be as-if randomly assigned, our approach allows investigators to determine if the instrument is at least significantly closer to being as-if randomly assigned than the exposure. In this scenario, an argument can be made that an IV analysis should still be conducted, even if the IV is not---strictly speaking---randomly assigned. To our knowledge, this more nuanced assessment is not provided by other falsification tests in the literature; we discuss this point further in Section \ref{ss:comparingInstrumentExposureBalance}.

%Besides these essential assumptions, point identification of the
%causal effect requires additional assumptions such as a deterministic
%\cite{Angrist:1996} or stochastic \cite{small2017instrumental}
%monotonicity assumption, or a no-interaction assumption
%\cite{Hernan:2006,wang2018bounded}. Here we will focus on core IV
%assumptions (2) and (3), since the relevance assumption (1) can be easily verified empirically (for example, by a
%regression of $D$ on $\bm X$ and $Z$).

\section{A Randomization Test for Instrument Validity} 
\label{s:randomizationTest}

As discussed in Section \ref{sec:intro}, previous works have tested for effective random assignment by comparing the IV and the exposure either in terms of covariate balance \cite{brookhart2007preference} or bias \cite{baiocchi2014instrumental,jackson2015toward,davies2015commentary,davies2017compare,zhao2018}. Covariate balance---also known as the covariate prevalence difference---is defined as the difference in covariate means across a binary measure of the IV. Meanwhile, in this context, bias is defined as the covariate prevalence difference divided by the treatment prevalence difference for the same covariate. Our test can incorporate covariate balance or bias. The intuition behind our test is that it compares the observed covariate balance or bias to the balance or bias we would expect from an as-if randomized experiment. Thus, our test proceeds by first establishing a standard for as-if randomization.

If Assumption (2) holds, we can write the conditional distribution of the instrument as $P(\bm Z | \bm D, \bm{Y}(\bm{D}), \bm{X}) = P(\bm{Z} | \bm{X})$, which is the conditional distribution of the instrument as a function of observed covariates. Thus, an instrument is as-if randomized if its distribution follows some $P(\bm{Z} | \bm{X})$, and our test will assess if indeed the distribution of the instrument follows some distribution $P(\bm{Z} | \bm{X})$. For example, in our application in Section \ref{s:application}, we will consider positing that the instrument follows \textit{complete randomization}, defined as:
\begin{align}
  P_{cr}(\bm{Z} = \bm{z} | \bm{X}) &= \begin{cases}
    {N \choose N_T}^{-1} &\mbox{ if $\sum_{i=1}^N z_i = N_T$} \\
    0 &\mbox{ otherwise.}
  \end{cases} \label{eqn:completeRandomization}
\end{align}
Under complete randomization, the instrument is uniformly randomized to all units conditional on the number of treated units $N_T$. In practice, the investigator can posit any assignment mechanism that can be written as a distribution $P(\bm{Z} | \bm{X})$, and our test can in turn assess the plausibility of that assignment mechanism. For example, if the units can be easily organized into blocks, then block randomization (instead of complete randomization) may be appropriate. See Imbens and Rubin\cite{imbens2015causal} (Chapter 4) for a full taxonomy of assignment mechanisms.

We will now outline a randomization test for the hypothesis $H_0: \bm{Z} \sim P(\bm{Z} | \bm{X})$, i.e., the hypothesis that the instrument is as-if randomized according to $P(\bm{Z} | \bm{X})$. Our test compares the observed covariate balance to the distribution of covariate balance we would expect if indeed $H_0$ were true. If the observed covariate balance is substantially outside this distribution, then this is evidence against $H_0$---i.e., this is evidence against the assumption that the instrument is as-if randomized. Alternatively, if the observed covariate balance falls substantially within this distribution, this is evidence that the instrument follows the distribution $P(\bm{Z} | \bm{X})$ specified by the investigator. The formal algorithm for our randomization test is: \\

\noindent\fbox{%
\parbox{\textwidth}{%
\textbf{Algorithm for $\alpha$-level Randomization Test for $H_0: \bm{Z} \sim P(\bm{Z} | \bm{X})$}
\begin{enumerate}
\setlength\itemsep{0em}
  \item Specify an assignment mechanism $P(\bm{Z} | \bm{X})$.
  \item Define a test statistic $t(\bm{Z}, \bm{X})$, which measures covariate balance or bias.
  \item Generate $M$ random draws $\bm{Z}^{(1)},\dots,\bm{Z}^{(M)} \sim P(\bm{Z} | \bm{X})$. Here, $M$ should be sufficiently large (e.g., 5,000 or more) such that the randomization distribution is well-approximated.
  \item Compute the following randomization-based $p$-value:
  \begin{align}
    p = \frac{1 + \sum_{m=1}^M \mathbb{I}(|t(\bm{z}^{(m)}, \bm{X})| \geq |t^{obs}|)}{M + 1}, \hspace{0.1 in} \text{where $t^{obs} \equiv t \left(\bm{Z}^{obs}, \bm{X} \right)$} \label{eqn:pvalue}
  \end{align}
  \item Reject $H_0$ if $p \leq \alpha$.
\end{enumerate}
  }
  } \\

A key advantage of our randomization test is that it is valid---i.e., the test controls the false rejection rate---and exact for the hypothesis that $H_0: \bm{Z} \sim P(\bm{Z} | \bm{X})$ for any assignment mechanism of interest \cite{branson2018my}. Moreover, the test does not rely on any parametric assumptions.

The procedure requires the investigator to select a test statistic $t(\bm{Z}, \bm{X})$, which can be any measure of covariate balance or bias. For example, the test statistic can be covariate-specific.  Two natural covariate-specific test statistics consistent with the literature are the covariate prevalence difference---which we also refer to as balance---or bias:
\begin{align}
	\underbrace{\overline{\mathbf{X}}_{Z=1} - \overline{\mathbf{X}}_{Z=0}}_{\text{covariate prevalence difference}}  \hspace{0.5 in} \text{or} \hspace{0.5 in} \underbrace{\frac{\overline{\mathbf{X}}_{Z=1} - \overline{\mathbf{X}}_{Z=0}}{\overline{\mathbf{D}}_{Z=1} - \overline{\mathbf{D}}_{Z=0}}}_{\text{IV bias}} \label{eqn:balanceBias}
\end{align}
where $\overline{\bm{X}}_{Z=z}$ is the vector of covariate means for units with $Z = z$, and $\overline{\mathbf{D}}_{Z=z}$ is analogously defined. Covariate-specific test statistics allow the analyst to assess to what degree the instrument is as-if randomized for each covariate. 

Alternatively, one can assess if the instrument is as-if randomized across a global measure of covariate balance, which is perhaps the biggest advantage of our test. One measure of global covariate balance is the Mahalanobis distance \cite{mahalanobis1936generalized}, defined as
  \begin{align}
    M_Z &= (\overline{\bm{X}}_{Z=1} - \overline{\bm{X}}_{Z=0})^T \left[ \text{cov}(\overline{\bm{X}}_{Z=1} - \overline{\bm{X}}_{Z=0}) \right]^{-1}(\overline{\bm{X}}_{Z=1} - \overline{\bm{X}}_{Z=0}) \label{eqn:md}
  \end{align}
Here we have defined the Mahalanobis distance in terms of balance. Holding the instrument strength $\overline{\bm{D}}_{Z = 1} - \overline{\bm{D}}_{Z = 0}$ fixed, this is equivalent to defining this quantity in terms of bias, because the Mahalanobis distance is affinely invariant, which implies:
 \begin{align*}
    \left(\frac{\overline{\mathbf{X}}_{Z=1} - \overline{\mathbf{X}}_{Z=0}}{\overline{\mathbf{D}}_{Z=1} - \overline{\mathbf{D}}_{Z=0}}\right)^T \left[ \text{cov} \left(\frac{\overline{\mathbf{X}}_{Z=1} - \overline{\mathbf{X}}_{Z=0}}{\overline{\mathbf{D}}_{Z=1} - \overline{\mathbf{D}}_{Z=0}}\right) \right]^{-1} \left(\frac{\overline{\mathbf{X}}_{Z=1} - \overline{\mathbf{X}}_{Z=0}}{\overline{\mathbf{D}}_{Z=1} - \overline{\mathbf{D}}_{Z=0}}\right) &= M_Z
    \end{align*}
Thus, a benefit of the Mahalanobis distance is that it acts as an omnibus analog to covariate balance as well as bias, while also standardizing by the covariance among the covariates instead of their marginal variances.

Regardless of the choice of test statistic, our test is always valid for the assumption that the instrument is as-if randomized. However, the choice of test statistic may also affect the power of our test, as is the case for any falsification test. We elaborate on this point in Section \ref{s:conclusion}.

Our test is identical to the randomization test proposed by Branson\cite{branson2018my}. However, Branson uses the above test to assess as-if randomization for exposure assignment in matched datasets, whereas we assess as-if randomization for instrument assignment in observational datasets. The original use of the test, however, suggests another extension. In our above formulation, we use the test to assess if the instrument is as-if randomized, and we can do the same for the exposure by replacing $\bm{Z}$ with $\bm{D}$ in our above formulation. Then, we can compare the conclusions of our test for $\bm{Z}$ and for $\bm{D}$, as we outline further in the next section. 

\subsection{Comparing Instrument and Exposure Balance or Bias} 
\label{ss:comparingInstrumentExposureBalance}

Thus far, we have outlined a randomization test that can assess if the instrument or the exposure are as-if randomized to the units under study. Table~\ref{tab:randomizationTestPossibilities} outlines the four possible scenarios that can occur when making these assessments. Case 1 provides clear evidence in favor of using the IV design, because the IV appears to be as-if randomized but the exposure does not. Case 2 is clear evidence against the IV design, because the exposure appears to be as-if randomized but the IV does not. Case 3 is an ambiguous but unlikely situation, where both the IV and the exposure appear to be as-if randomized. Case 4 is probably the most likely scenario in an observational study, where neither the IV nor the exposure appear to be as-if randomized. One dogmatic conclusion that one could draw from Case 4 is that no analysis should be conducted. A more pragmatic view is that, even if neither the instrument nor the exposure appear to be as-if randomized, it may still be desirable to conduct an IV analysis if the instrument appears to at least be significantly closer to as-if random than the exposure. This pragmatic view is consistent with the notion that natural experiments are useful if they at least reduce bias relative to cases where treatments are self-selected.\cite{rosenbaum2010design,dunning2012natural,Rosenbaum:1999}

\begin{table}
\centering
  \begin{tabular}{c|c|c||c}
  \toprule
   & \textbf{Exposure} ($\bm{D}$) & \textbf{Instrument} ($\bm{Z}$) & \textbf{Recommendation} \\
   \midrule
    \textbf{Case 1} & Reject $H_0$ & Fail to Reject $H_0$ & Use IV analysis \\
    \textbf{Case 2} & Fail to Reject $H_0$ & Reject $H_0$ & Reject IV analysis \\
    \textbf{Case 3} & Fail to Reject $H_0$ & Fail to Reject $H_0$ & Use IV analysis or exposure analysis \\
    \textbf{Case 4} & Reject $H_0$ & Reject $H_0$ & Compare balance or bias of $\bm{D}$ and $\bm{Z}$ \\
    \bottomrule
  \end{tabular}
  \caption{The four possible cases (and our corresponding recommendation) when using our randomization test to assess the hypothesis of as-if randomization ($H_0$) for $\bm{D}$ and $\bm{Z}$.}
  \label{tab:randomizationTestPossibilities}
\end{table}

Accordingly, we now outline how to test whether the instrument is significantly closer to being as-if randomized than the exposure. For this test, we must consider alternative assignment mechanisms that the exposure or instrument may follow, because, under Case 4 in Table \ref{tab:randomizationTestPossibilities}, we have rejected that they are completely randomized to the units under study. One mechanism to posit is that exposure and instrument assignment follow \textit{Bernoulli trials}:
\begin{align}
  P_{bt}(\bm{D} = \bm{d}| \bm{X}) &= \prod_{i=1}^N e_D(\bm{x}_i)^{d_i} [1 - e_d(\bm{x}_i)]^{1-d_i} \label{eqn:exposureBT} \\
  P_{bt}(\bm{Z} = \bm{z} | \bm{X}) &= \prod_{i=1}^N e_Z(\bm{x}_i)^{z_i} [1 - e_z(\bm{x}_i)]^{1-z_i} \label{eqn:instrumentBT}
\end{align}
where $e_D(\bm{x}_i) \equiv P(\bm{D}_i = 1 | \bm{x}_i)$ is the $i$th propensity score for receiving the exposure, and $e_Z(\bm{x}_i)$ is the $i$th propensity score analogously defined for the instrument. This is a natural mechanism to posit in observational studies, because such a mechanism is implicitly assumed when conducting propensity score analyses, which are widely used for observational studies\cite{rosenbaum2002covariance,rubin2007design,rubin2008objective,branson2018randomization}. These propensity scores are not known, but they can be easily estimated via logistic regression. 

After the propensity scores are estimated, we can generate many hypothetical randomizations according to $P_{bt}(\bm{D} | \bm{X})$ and $P_{bt}(\bm{Z} | \bm{X})$ via biased coin flips, and then compute the balance or bias across these hypothetical randomizations; this will measure the balance or bias we would expect from the instrument assignment mechanism and the exposure assignment mechanism. This can be done by the following procedure:
\begin{enumerate}
  \item Estimate the exposure propensity scores $e_D(\bm{x})$ and the instrument propensity scores $e_Z(\bm{x})$.
  \item Generate many hypothetical exposure assignments and instrument assignments:
  \begin{align}
    \bm{d}^{(1)}, \dots, \bm{d}^{(M)} &\sim \hat{P}_{bt}(\bm{D} | \bm{X}) \\
    \bm{z}^{(1)}, \dots, \bm{z}^{(M)} &\sim \hat{P}_{bt}(\bm{Z} | \bm{X})
  \end{align}
  where $\hat{P}_{bt}(\bm{D} | \bm{X})$ is $P_{bt}(\bm{D} | \bm{X})$ in (\ref{eqn:exposureBT}), but using the estimated $\hat{e}_D(\bm{x})$ from Step 1; $\hat{P}_{bt}(\bm{Z} | \bm{X})$ is analogously defined.
  \item Compute the test statistic (balance or bias) $t(\bm{D}, \bm{X})$ and $t(\bm{Z}, \bm{X})$ across the hypothetical exposure assignments and instrument assignments, respectively, from Step 2.
  \item Compare the randomization distributions $(t(\bm{d}^{(1)}, \bm{X}), \dots, t(\bm{d}^{(M)}, \bm{X}))$ and $(t(\bm{z}^{(1)}, \bm{X}), \dots, t(\bm{z}^{(M)}, \bm{X}))$.
\end{enumerate}
If the randomization distributions are non-overlapping, this implies that balance or bias across the exposure and across the instrument significantly differ. Furthermore, both of these randomization distributions can be compared to the corresponding randomization distribution under \textit{complete randomization} defined in (\ref{eqn:completeRandomization}). For this comparison, global measures of balance or bias---such as the Mahalanobis distance---are particularly useful, because they place gold standard randomized assignment, exposure assignment, and instrument assignment on a univariate scale, making comparisons straightforward. As we will see in the next section, it is especially helpful to plot the randomization distribution across these three assignment mechanisms, because this aids in assessing if the instrument is significantly closer to being as-if randomized than the exposure. Next, we demonstrate the proposed methods in a case study.

\section{Application: Is Bed Availability a Valid Instrument for ICU Admission?} 
\label{s:application}

We use data from (SPOT)light, a prospective cohort study of deteriorating ward patients referred to critical care in 48 National Health Service (NHS) hospitals between 1 November 2010 and 31 December 2011 \cite{harrisicu2018}. The cohort includes 13,011 patients. The primary clinical question of interest was whether transfer to the ICU reduced in-hospital mortality. Bias from confounding by indication may be a threat if the reasons for transfer to the ICU are not fully recorded in the data. Threat of this bias motivated the use of an IV analysis. During data collection, ICU bed availability at the time of assessment for ICU admission was measured for use as an instrument for ICU transfer \cite{harrisicu2018,keele2018icubeds, keeleicumatch2019,kennedy2018survivor}. That is, if there are few ICU beds available when the patient is assessed, this discourages ICU transfer. In the data, ICU bed availability ranges from 0 to 19 with a median of 4 and an IQR of 4. Following an earlier analysis, we use a binary measure of the instrument where 1 indicates that fewer than 4 ICU beds were available \cite{keele2018icubeds}.

The exposure is defined as transfer to and receipt of care in the ICU. Measured confounders include age, gender, septic diagnosis (0/1), and peri-arrest (0/1). Physiological data were also collected to create three measures of patient risk severity. The data also record patients' existing level of care at assessment and the recommended level of care after assessment, defined using the UK Critical Care Minimum Dataset (CCMDS) levels of care. These levels are 0 and 1 for normal ward care, 2 for care within a high-dependency unit, and 3 for care within the ICU.

\begin{figure}
  \begin{subfigure}[t]{0.51\textwidth}
    \includegraphics[scale=0.45]{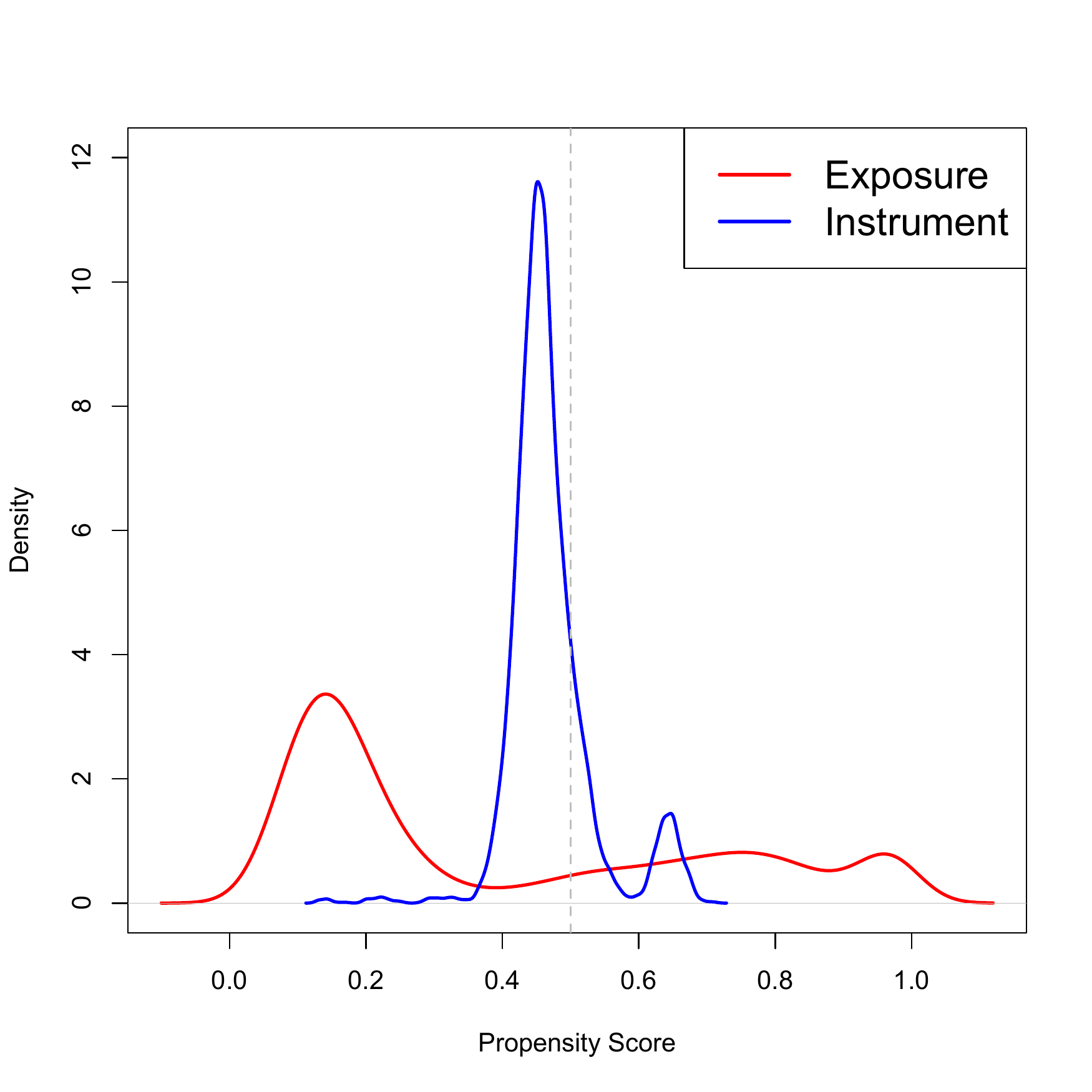}
    \caption{Propensity score distributions.}
    \label{fig:ps}
  \end{subfigure}
  ~
  \begin{subfigure}[t]{0.51\textwidth}
    \includegraphics[scale=0.45]{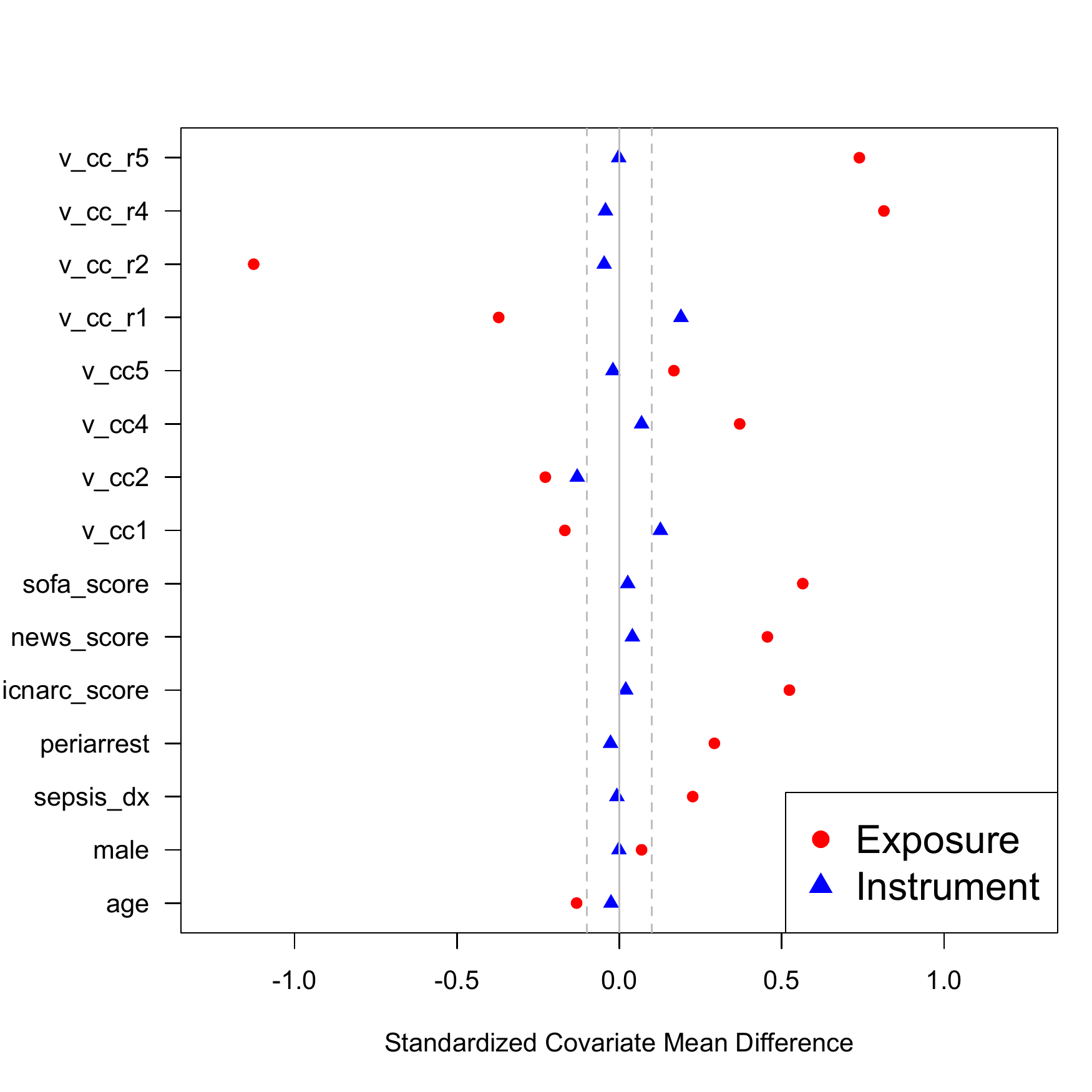}
    \caption{SCMDs. Dotted vertical lines: 0.1.}
    \label{fig:covMeanDiffs}
  \end{subfigure}
  \caption{Propensity scores and SCMDs across the exposure and the instrument.}
  \label{fig:psAndCovMeanDiffs}
\end{figure}

Figure \ref{fig:ps} shows the distribution of the propensity scores (estimated using logistic regression) across the exposure and across the instrument. The IV propensity scores are much more closely concentrated around moderate values, indicating better balance. Meanwhile, Figure \ref{fig:covMeanDiffs} shows the standardized covariate mean differences (SCMDs)---in other words, the standardized prevalence difference defined in (\ref{eqn:balanceBias})---across the instrument and across the exposure. SCMDs across the instrument are more closely concentrated around zero---and many are below the 0.1 rule-of-thumb that is commonly employed in the literature \cite{normand2001validating,austin2009some,zubizarreta2012using,resa2016evaluation}---again indicating better balance for the instrument. Figures \ref{fig:ps} and \ref{fig:covMeanDiffs} are useful diagnostics and are similar in spirit to current practice, but they do not tell us if the instrument balance is better than exposure balance to a statistically significant degree, or if the instrument can be assumed to be as-if randomized. 

Next, we apply the randomization test from Section \ref{s:randomizationTest}. First, we implement the test for each covariate using the SCMDs as the test statistic. Specifically, we implemented the randomization test by permuting the instrument indicator 10,000 times and computing the SCMDs for each permutation. The distribution of the resulting 10,000 SCMDs represent the balance we would expect from a completely randomized experiment. If the observed balance is significantly outside this distribution, then this is evidence against as-if randomization.

\begin{figure}
\centering
    \includegraphics[scale=0.45]{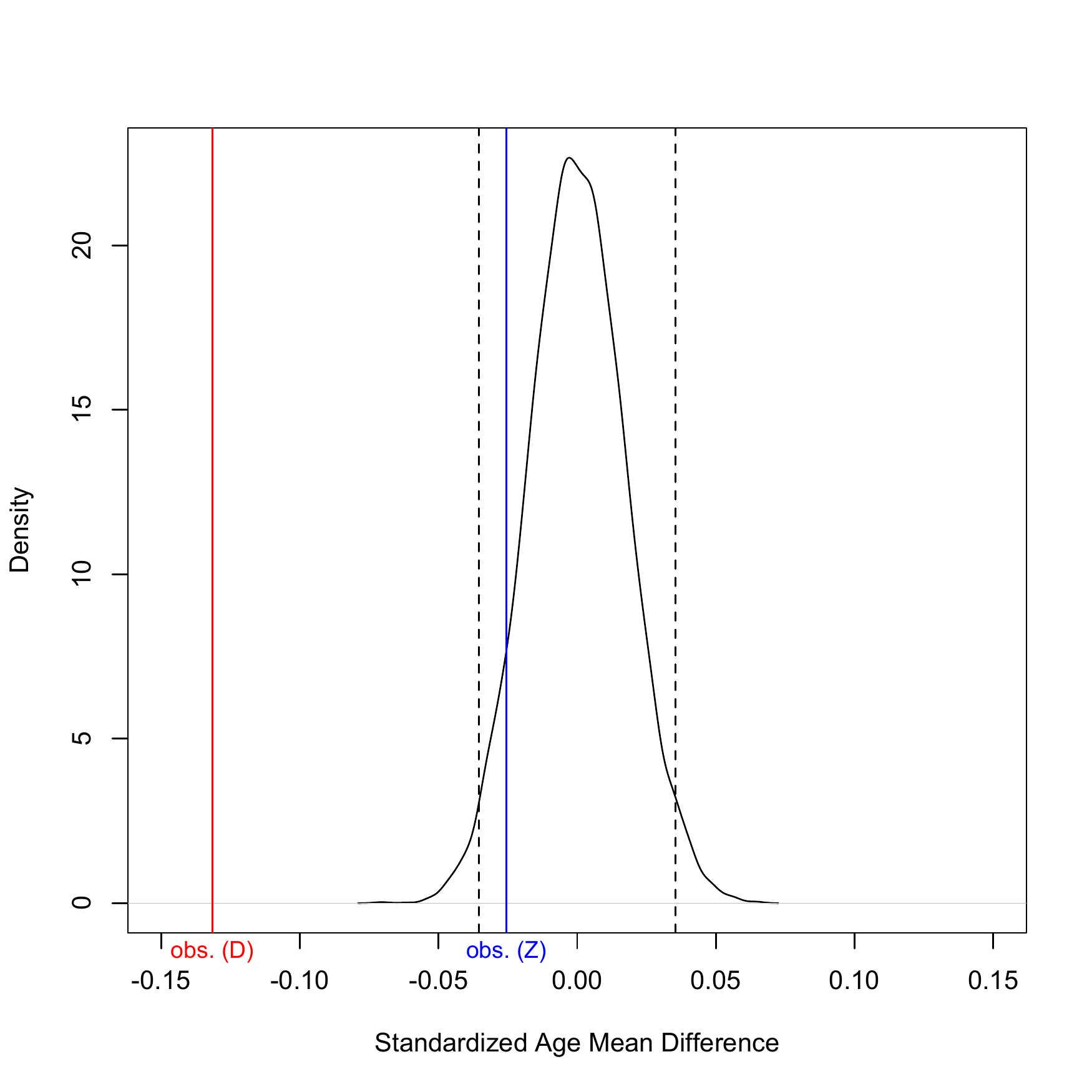}
    \caption{Permutation distribution of the SCMD for age. Dotted vertical lines denote the 2.5\% and 97.5\% permutation quantiles. Solid vertical lines denote observed balance for the instrument (``obs. (Z)'') and the exposure (``obs. (D)'').}
    \label{fig:covMeanDiffsAgePlot}
\end{figure}

\begin{figure}
  \begin{subfigure}[t]{0.51\textwidth}
    \includegraphics[scale=0.45]{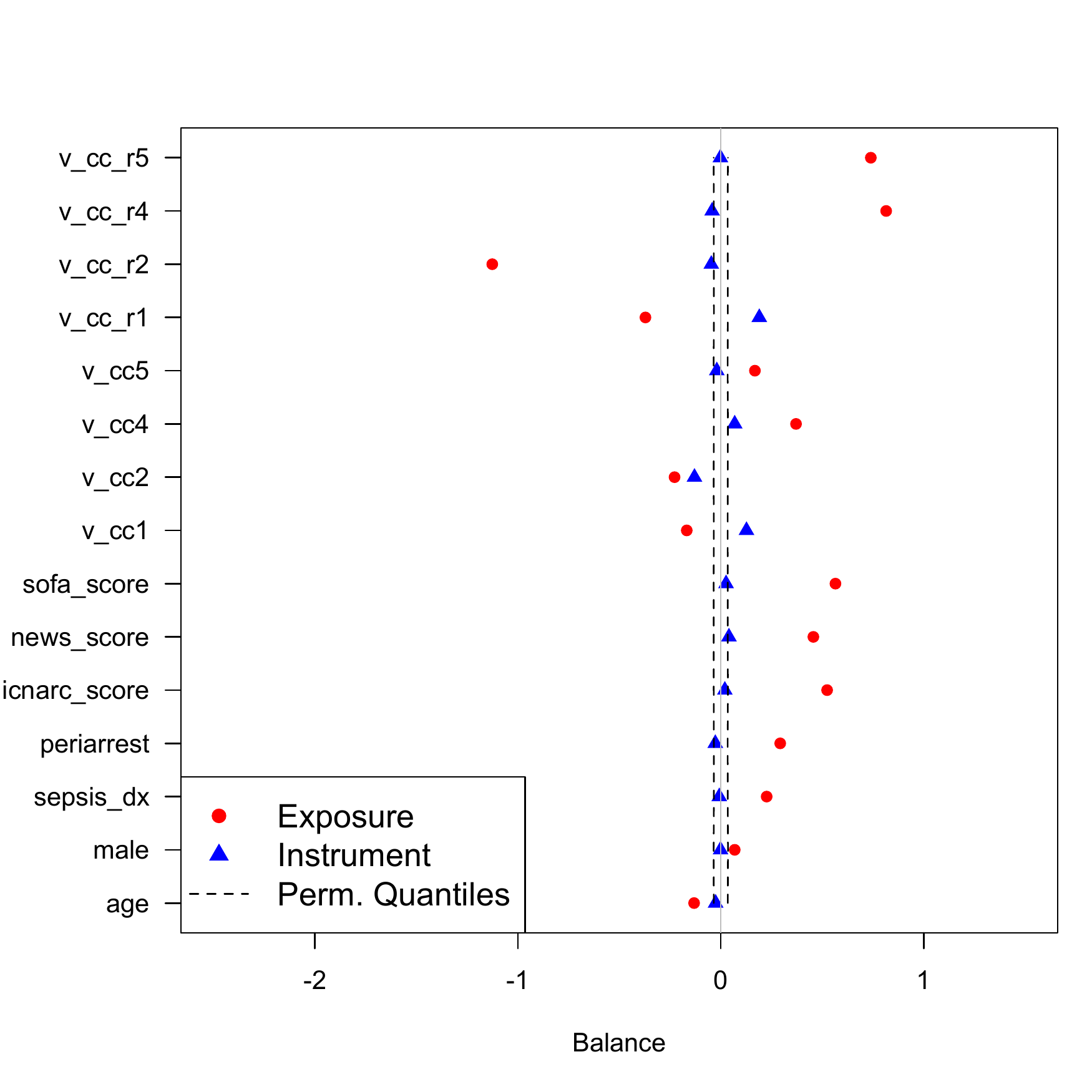}
    \caption{Permutation quantiles for balance.}
    \label{fig:balancePlot}
  \end{subfigure}
  ~
  \begin{subfigure}[t]{0.51\textwidth}
    \includegraphics[scale=0.45]{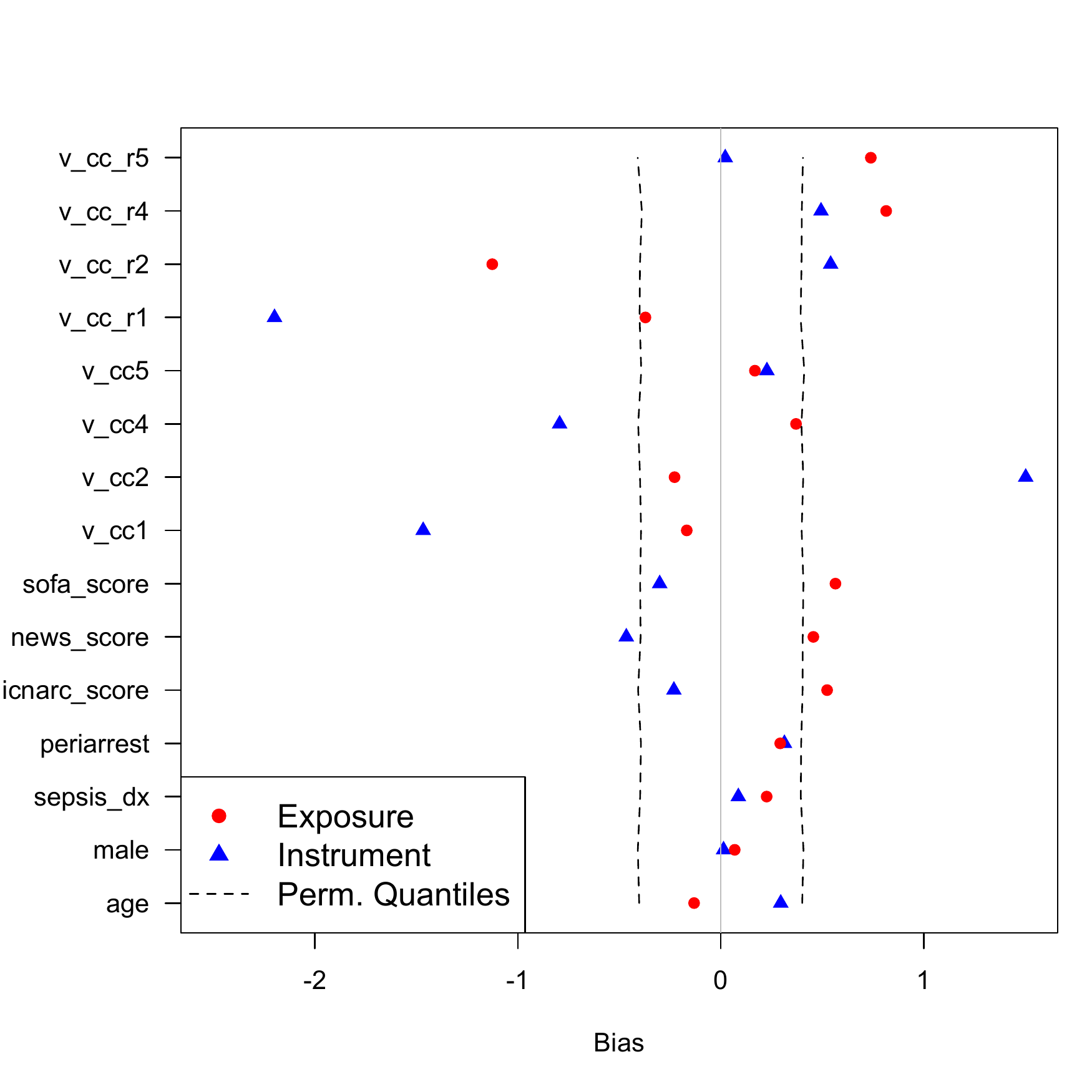}
    \caption{Permutation quantiles for bias.}
    \label{fig:biasPlot}
  \end{subfigure}
  \caption{2.5\% (left dotted line) and 97.5.\% (right dotted line) permutation quantiles for balance as measured by SCMDs (left plot) and bias (right plot) across all covariates.}
  \label{fig:balanceBiasPlot}
\end{figure}

For example, Figure \ref{fig:covMeanDiffsAgePlot} shows the permutation distribution of the SCMD for the age covariate. The observed SCMD for the instrument is within the 2.5\% and 97.5\% quantiles of this permutation distribution, meaning that our test concludes that the instrument is as-if randomized for this covariate with 95\% confidence (the randomization test $p$-value is 0.15). Furthermore, the observed SCMD for the exposure is not within these quantiles, indicating that the exposure is not as-if randomized for this covariate. Meanwhile, Figure \ref{fig:balancePlot} shows the SCMDs for the other covariates and their corresponding permutation quantiles. All of the observed SCMDs across the exposure are outside these quantiles, and thus as-if randomization is not tenable for the exposure. However, the observed SCMDs across the instrument are within these quantiles for some of the covariates but not others, making it unclear whether as-if randomization can be assumed for the instrument.

The situation is even more ambiguous if we look at bias instead of balance, which has been recently recommended in the literature\cite{baiocchi2014instrumental,jackson2015toward,davies2015commentary,davies2017compare,zhao2018}. Figure~\ref{fig:biasPlot} shows the instrument and exposure bias for each covariate---as defined in (\ref{eqn:balanceBias})---and their corresponding permutation quantiles. Note that exposure bias is equal to exposure balance, while instrument bias is instrument balance inflated by the strength of the IV. Thus, the red dots in Figures \ref{fig:balancePlot} and \ref{fig:biasPlot} representing exposure balance and bias are identical, while the blue triangles representing instrument bias in Figure \ref{fig:biasPlot} are inflated versions of those in Figure \ref{fig:balancePlot}. Now it is less clear if the IV is preferable over the exposure in terms of bias. 

\begin{figure}
  \centering
  \includegraphics[scale=0.75]{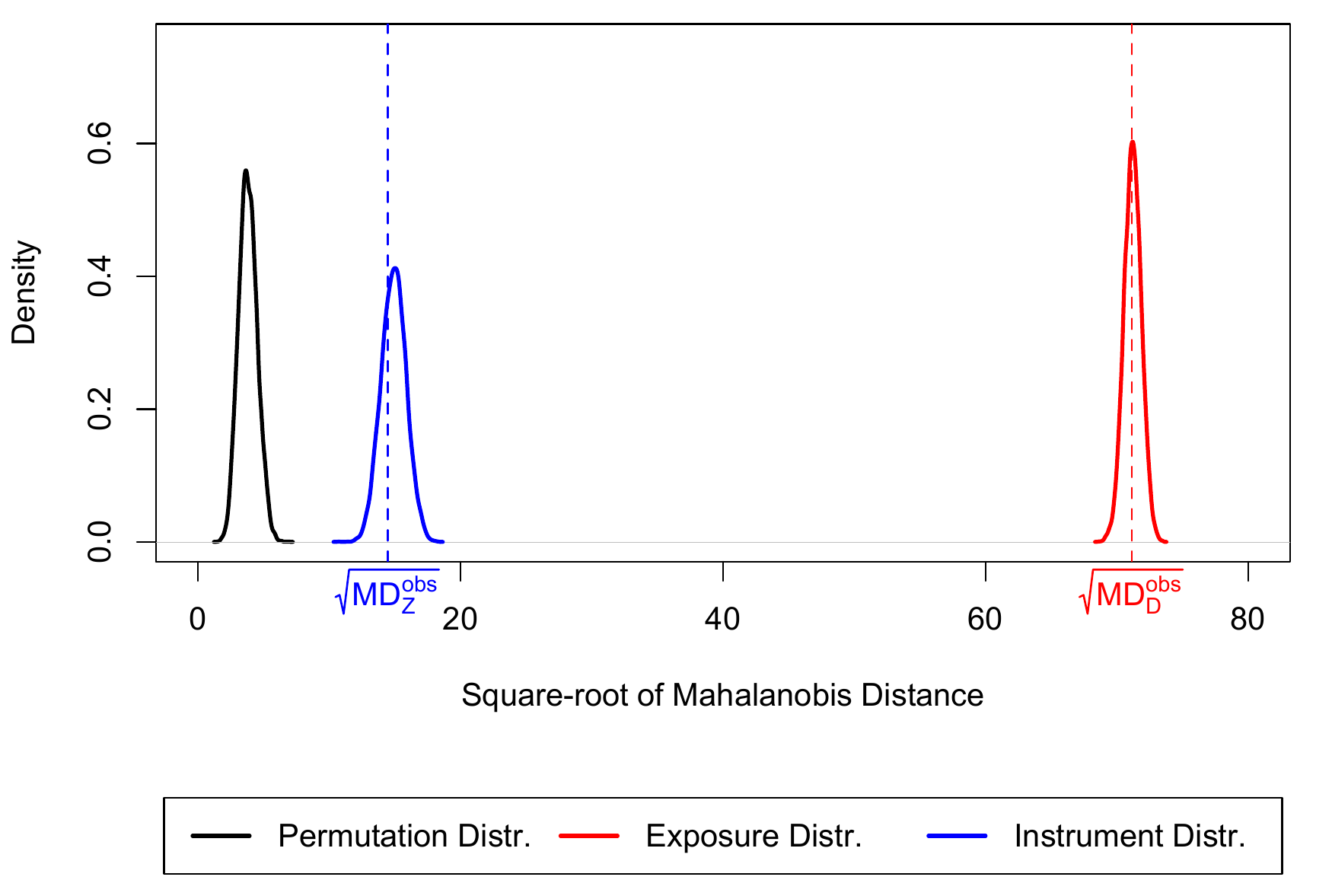}
  \caption{Distribution of the square-root of the Mahalanobis distance across 10,000 permutations of the instrument (in black), 10,000 biased-coin randomizations using instrument propensity scores (in blue), and 10,000 biased-coin randomizations using exposure propensity scores (in red). The observed square-root of the Mahalanobis distance for the instrument ($\sqrt{\text{MD}_{Z}^{\text{obs}}}$) and exposure ($\sqrt{\text{MD}_{D}^{\text{obs}}}$) are marked by blue and red dotted vertical lines, respectively.}
  \label{fig:mdPlot}
\end{figure}

The ambiguities in covariate-by-covariate diagnostics make global measures of balance or bias useful. Next, we implement our test using the square-root of the Mahalanobis distance---i.e., the square-root of (\ref{eqn:md})---which acts as a global measure of balance as well as bias. We again permute the instrument indicator 10,000 times and compute the square-root of the Mahalanobis distance for each permutation. The resulting permutation distribution is shown in black in Figure \ref{fig:mdPlot}. The observed square-root of the Mahalanobis distance for the instrument ($\sqrt{\text{MD}_{Z}^{\text{obs}}}$) and exposure ($\sqrt{\text{MD}_{D}^{\text{obs}}}$)---marked by vertical lines---are both outside this distribution. Therefore, we conclude that neither the instrument nor the exposure are completely randomized.

Because we have rejected the null hypothesis that instrument and exposure assignment follow complete randomization, we must posit alternative assignment mechanisms that the instrument and exposure may follow. Next, we assume that the instrument and exposure assignment mechanisms follow Bernoulli trials, denoted by (\ref{eqn:exposureBT}) and (\ref{eqn:instrumentBT}), and we will estimate the propensity scores for these assignment mechanisms using logistic regression, as we did in Figure \ref{fig:ps}. Then, we generate 10,000 random draws from these assignment mechanisms and compute the square-root of the Mahalanobis distance for each draw. The resulting 10,000 Mahalanobis distances represent the balance or bias we would expect from a Bernoulli trial experiment using the estimated propensity scores. The randomization distributions under instrument assignment and exposure assignment are shown in blue and red, respectively, in Figure \ref{fig:mdPlot}. Two observations can be made: (1) these distributions are non-overlapping, and (2) the instrument distribution is substantially closer to the as-if randomization distribution than the exposure distribution. Thus, while we cannot conclude that the instrument is as-if randomized, we can at least be confident that the IV is substantially closer to being as-if randomized than the exposure. This observation can be used to justify conducting an IV analysis on these data, which has been done in several recent works.\cite{harrisicu2018,keele2018icubeds, keeleicumatch2019,kennedy2018survivor}

\section{Conclusions}
\label{s:conclusion}

Much recent work has developed statistical tools to help investigators judge whether a posited instrument is valid. Here, we add to this literature by developing a randomization test to probe the assumption that the IV is effectively randomly assigned. Our test compares instrument and exposure balance or bias to the balance or bias we would expect from a randomized experiment. This allows us to assess if the instrument is as-if randomized as well as assess if the instrument is significantly closer to being as-if randomized than the exposure. This test has several key advantages. First, our randomization test is an exact and valid test for the effective random assignment assumption, and it does not rely on any parametric assumptions. Second, our test can provide covariate-by-covariate results as well as a global test, and it can incorporate measures of balance as well as bias. Third, our test uses what we would expect from a randomized experiment as a clear benchmark. Finally, the results from the test can be summarized in a single graphical display and are easily interpretable. We demonstrated our approach with a recent application that uses bed availability in the ICU as an instrument for admission to the ICU. In that application, we found that neither the instrument nor the exposure is effectively randomized; however, we also found that the instrument is substantially closer to being as-if randomized than the exposure. 

There are several limitations to our test, and addressing these limitations would be an interesting avenue for future work. First, we have focused on the case where the exposure and the instrument are binary, and it would be helpful to extend our test to cases where the exposure and/or the instrument are multi-valued. Furthermore, while our test provides a valid way to assess as-if randomization for an instrument in observational studies, the test may be under-powered. This is not a limitation of our randomization test specifically but of falsification tests in general. For example, most tests in the literature focus on comparing instrument and exposure balance in terms of the covariates' first moments---and we have done the same here---but these tests will have little power in detecting differences in other functions of the covariates. Of course, our test allows for nonlinear functions of the covariates, but these functions must be specified by the investigator. Future work could consider methods that are both valid and have optimal power in detecting violations of the effective random assignment assumption.

\clearpage
\singlespacing

\bibliographystyle{ieeetr}
\bibliography{bransonKeeleIVBib}

\begin{thebibliography}{10}

\bibitem{imbens1997bayesian}
G.~W. Imbens and D.~B. Rubin, ``Bayesian inference for causal effects in
  randomized experiments with noncompliance,'' {\em The Annals of Statistics},
  pp.~305--327, 1997.

\bibitem{dunn2005estimating}
G.~Dunn, M.~Maracy, and B.~Tomenson, ``Estimating treatment effects from
  randomized clinical trials with noncompliance and loss to follow-up: the role
  of instrumental variable methods,'' {\em Statistical Methods in Medical
  Research}, vol.~14, no.~4, pp.~369--395, 2005.

\bibitem{angrist1996identification}
J.~D. Angrist, G.~W. Imbens, and D.~B. Rubin, ``Identification of causal
  effects using instrumental variables,'' {\em Journal of the American
  Statistical Association}, vol.~91, no.~434, pp.~444--455, 1996.

\bibitem{baiocchi2014instrumental}
M.~Baiocchi, J.~Cheng, and D.~S. Small, ``Instrumental variable methods for
  causal inference,'' {\em Statistics in Medicine}, vol.~33, no.~13,
  pp.~2297--2340, 2014.

\bibitem{baiocchi2012near}
M.~Baiocchi, D.~S. Small, L.~Yang, D.~Polsky, and P.~W. Groeneveld, ``Near/far
  matching: a study design approach to instrumental variables,'' {\em Health
  Services and Outcomes Research Methodology}, vol.~12, no.~4, pp.~237--253,
  2012.

\bibitem{swanson2013commentary}
S.~A. Swanson and M.~A. Hern{\'a}n, ``Commentary: {H}ow to report instrumental
  variable analyses (suggestions welcome),'' {\em Epidemiology}, vol.~24,
  no.~3, pp.~370--374, 2013.

\bibitem{Yang:2013}
F.~Yang, J.~Zubizaretta, D.~S. Small, S.~Lorch, and P.~Rosenbaum, ``Dissonant
  conclusions when testing the validity of an instrumental variable,'' {\em The
  American Statistician}, vol.~68, no.~4, pp.~253--263, 2014.

\bibitem{glymour2012credible}
M.~M. Glymour, E.~J. Tchetgen~Tchetgen, and J.~M. Robins, ``Credible mendelian
  randomization studies: {A}pproaches for evaluating the instrumental variable
  assumptions,'' {\em American Journal of Epidemiology}, vol.~175, no.~4,
  pp.~332--339, 2012.

\bibitem{kang2013causal}
H.~Kang, B.~Kreuels, O.~Adjei, R.~Krumkamp, J.~May, and D.~S. Small, ``The
  causal effect of malaria on stunting: {A} mendelian randomization and
  matching approach,'' {\em International Journal of Epidemiology}, vol.~42,
  no.~5, pp.~1390--1398, 2013.

\bibitem{pizer2016falsification}
S.~D. Pizer, ``Falsification testing of instrumental variables methods for
  comparative effectiveness research,'' {\em Health Services Research},
  vol.~51, no.~2, pp.~790--811, 2016.

\bibitem{keeleexrest2018}
L.~J. Keele, Q.~Zhao, R.~R. Kelz, and D.~S. Small, ``Falsification tests for
  instrumental variable designs with an application to tendency to operate.,''
  {\em Medical Care}, vol.~57, no.~2, pp.~167--171, 2019.

\bibitem{brookhart2007preference}
M.~A. Brookhart and S.~Schneeweiss, ``Preference-based instrumental variable
  methods for the estimation of treatment effects: assessing validity and
  interpreting results,'' {\em The International Journal of Biostatistics},
  vol.~3, no.~1, 2007.

\bibitem{jackson2015toward}
J.~W. Jackson and S.~A. Swanson, ``Toward a clearer portrayal of confounding
  bias in instrumental variable applications,'' {\em Epidemiology (Cambridge,
  Mass.)}, vol.~26, no.~4, p.~498, 2015.

\bibitem{davies2015commentary}
N.~M. Davies, ``Commentary: {A}n even clearer portrait of bias in observational
  studies?,'' {\em Epidemiology (Cambridge, Mass.)}, vol.~26, no.~4, p.~505,
  2015.

\bibitem{davies2017compare}
N.~M. Davies, K.~H. Thomas, A.~E. Taylor, G.~M. Taylor, R.~M. Martin, M.~R.
  Munaf{\`o}, and F.~Windmeijer, ``How to compare instrumental variable and
  conventional regression analyses using negative controls and bias plots,''
  {\em International Journal of Epidemiology}, vol.~46, no.~6, pp.~2067--2077,
  2017.

\bibitem{zhao2018}
Q.~Zhao and D.~S. Small, ``Graphical diagnosis of confounding bias in
  instrumental variables analysis,'' {\em Epidemiology}, vol.~29, no.~4,
  pp.~e29--e31, 2018.

\bibitem{rubin1980randomization}
D.~B. Rubin, ``Randomization analysis of experimental data: The {F}isher
  randomization test (comment),'' {\em Journal of the American Statistical
  Association}, vol.~75, no.~371, pp.~591--593, 1980.

\bibitem{imbens2015causal}
G.~W. Imbens and D.~B. Rubin, {\em Causal inference in {S}tatistics, {S}ocial,
  and {B}iomedical {S}ciences}.
\newblock Cambridge University Press, 2015.

\bibitem{branson2018my}
Z.~Branson, ``Is my matched dataset as-if randomized, more, or less? {U}nifying
  the design and analysis of observational studies,'' {\em arXiv preprint
  arXiv:1804.08760}, 2018.

\bibitem{mahalanobis1936generalized}
P.~C. Mahalanobis, ``On the generalized distance in statistics,'' National
  Institute of Science of India, 1936.

\bibitem{rosenbaum2010design}
P.~R. Rosenbaum, {\em Design of {O}bservational {S}tudies}, vol.~10.
\newblock Springer, 2010.

\bibitem{dunning2012natural}
T.~Dunning, {\em Natural {E}xperiments in the {S}ocial {S}ciences: {A}
  {D}esign-based {A}pproach}.
\newblock Cambridge University Press, 2012.

\bibitem{Rosenbaum:1999}
P.~R. Rosenbaum, ``Choice as an alternative to control in observational
  studies,'' {\em Statistical Science}, vol.~14, no.~3, pp.~259--304, 1999.

\bibitem{rosenbaum2002covariance}
P.~R. Rosenbaum, ``Covariance adjustment in randomized experiments and
  observational studies,'' {\em Statistical Science}, vol.~17, no.~3,
  pp.~286--327, 2002.

\bibitem{rubin2007design}
D.~B. Rubin, ``The design versus the analysis of observational studies for
  causal effects: {P}arallels with the design of randomized trials,'' {\em
  Statistics in Medicine}, vol.~26, no.~1, pp.~20--36, 2007.

\bibitem{rubin2008objective}
D.~B. Rubin, ``For objective causal inference, design trumps analysis,'' {\em
  The Annals of Applied Statistics}, vol.~2, no.~3, pp.~808--840, 2008.

\bibitem{branson2018randomization}
Z.~Branson and M.-A. Bind, ``Randomization-based inference for bernoulli trial
  experiments and implications for observational studies,'' {\em Statistical
  Methods in Medical Research}, p.~0962280218756689, 2018.

\bibitem{harrisicu2018}
S.~Harris, M.~Singer, C.~S. C, R.~Grieve, D.~Harrison, and K.~Rowan, ``Impact
  on mortality of prompt admission to critical care for deteriorating ward
  patients: {A}n instrumental variable analysis using critical care bed
  strain,'' {\em Intensive Care Medicine}, vol.~44, no.~5, pp.~606--615, 2018.

\bibitem{keele2018icubeds}
L.~J. Keele, S.~Harris, and R.~D. Grieve, ``Does transfer to intensive care
  units reduce mortality? {A} comparison of an instrumental variables design to
  risk adjustment.,'' {\em Medical Care}, vol.~In press., 2019.

\bibitem{keeleicumatch2019}
L.~J. Keele, S.~Harris, S.~Pimentel, and R.~Grieve, ``Stronger instruments and
  refined covariate balance in an observational study of the effectiveness of
  prompt admission to the {ICU},'' {\em Journal of The Royal Statistical
  Society, Series A}, vol.~In press., 2019.

\bibitem{kennedy2018survivor}
E.~H. Kennedy, S.~Harris, and L.~J. Keele, ``Survivor-complier effects in the
  presence of selection on treatment, with application to a study of prompt
  {ICU} admission,'' {\em Journal of the American Statistical Association},
  pp.~1--12, 2018.

\bibitem{normand2001validating}
S.-L.~T. Normand, M.~B. Landrum, E.~Guadagnoli, J.~Z. Ayanian, T.~J. Ryan,
  P.~D. Cleary, and B.~J. McNeil, ``Validating recommendations for coronary
  angiography following acute myocardial infarction in the elderly: {A} matched
  analysis using propensity scores,'' {\em Journal of Clinical Epidemiology},
  vol.~54, no.~4, pp.~387--398, 2001.

\bibitem{austin2009some}
P.~C. Austin, ``Some methods of propensity-score matching had superior
  performance to others: {R}esults of an empirical investigation and monte
  carlo simulations,'' {\em Biometrical Journal}, vol.~51, no.~1, pp.~171--184,
  2009.

\bibitem{zubizarreta2012using}
J.~R. Zubizarreta, ``Using mixed integer programming for matching in an
  observational study of kidney failure after surgery,'' {\em Journal of the
  American Statistical Association}, vol.~107, no.~500, pp.~1360--1371, 2012.

\bibitem{resa2016evaluation}
M.~Resa and J.~R. Zubizarreta, ``Evaluation of subset matching methods and
  forms of covariate balance,'' {\em Statistics in Medicine}, vol.~35, no.~27,
  pp.~4961--4979, 2016.

\end{thebibliography}

\end{document}